# *Orbital selective switching of ferromagnetism in an oxide quasi two-dimensional electron gas*

R. Di Capua[1,2], M. Verma[3], M. Radovic[4], V.N. Strocov[4], C. Piamonteze[4], E. B. Guedes[4], N. Plumb[4], Yu Chen[2], M. D'Antuono[1,2], G.M. De Luca[1,2], E. Di Gennaro[1,2], D. Stornaiuolo[1,2], D. Preziosi[5], B. Jouault[6], F. Miletto Granozio[2], A. Sambri[2], R. Pentcheva[3], G. Ghiringhelli[7,8] and M. Salluzzo[2*]

[1] *Dipartimento di Fisica "E. Pancini", Università di Napoli "Federico II", Complesso Monte Sant'Angelo via Cinthia, I-80126 Napoli, Italy.*
[2] *CNR-SPIN, Complesso Monte Sant'Angelo via Cinthia, I-80126 Napoli, Italy.*
[3] *Department of Physics and Center for Nanointegration, University Duisburg-Essen Lotharstr. 1, D-47057 Duisburg, Germany.*
[4] *Photon Science Division, Paul Scherrer Institut, CH-5232 Villigen PSI, Switzerland.*
[5] *Université de Strasbourg, CNRS, IPCMS UMR 7504, 67034 Strasbourg, France.*
[6] *Laboratoire Charles Coulomb, UMR 5221, CNRS, Université de Montpellier, F-34095 Montpellier, France*
[7] *Dipartimento di Fisica Politecnico di Milano, Piazza Leonardo da Vinci 32, I-20133 Milano, Italy.*
[8] *CNR-SPIN, Politecnico di Milano, Piazza Leonardo da Vinci 32, I-20133 Milano, Italy.*

* marco.salluzzo@spin.cnr.it

**ABSTRACT**

**Multi-orbital physics in quasi-two-dimensional electron gases (q2DEGs) triggers unique phenomena not observed in bulk materials, such as unconventional superconductivity and magnetism. Here, we investigate the mechanism of orbital selective switching of the spin-polarization in the oxide q2DEG formed at the (001) interface between the $LaAlO_3$, $EuTiO_3$ and $SrTiO_3$ band insulators. By using density functional theory calculations, transport, magnetic and x-ray spectroscopy measurements, we find that the filling of titanium-bands with *$3d_{xz,yz}$* orbital character in the $EuTiO_3$ layer and at the interface with $SrTiO_3$ induces an antiferromagnetic to ferromagnetic switching of the exchange interaction between Eu-*$4f^7$* magnetic moments. The results explain the observation of the carrier density dependent ferromagnetic correlations and anomalous Hall effect in this q2DEG, and demonstrate how combined theoretical and experimental approaches can lead to a deeper understanding of novel electronic phases and serve as a guide for the materials design for advanced electronic applications.**

**Introduction**

Since the discovery of a quasi-two-dimensional electron gas (q2DEG) at the interface between the $LaAlO_3$ (LAO) and $SrTiO_3$ (STO) band insulators [1], studies on oxide surfaces and interfaces uncovered an intriguing and rich physics, boosting the expectations for all-oxide electronics. The LAO/STO q2DEG shows remarkable properties, including electric field effect induced insulator-to-metal transition at room temperature [2], gate-tunable Rashba-like spin-orbit coupling [3, 4], superconductivity [5-7], and magnetism [8]. From these studies, a fundamental role of the multi-orbital nature of the carriers in the electronic properties of the oxide q2DEGs clearly emerged.

The possibility to induce ferromagnetic (FM) correlations at the interface between non-magnetic oxides, combined with the large spin to charge conversion efficiency of oxide q2DEGs [9], paves the way to applications in spintronics. However, the reported magnetism at the bare LAO/STO interface is believed to be related to defects and oxygen vacancies [10] more than intrinsic coupling of electronic states.

Recently, it has been shown that a feasible method to induce a spin-polarization in oxide q2DEG is the introduction of a thin magnetic layer between LAO and STO, like $EuTiO_3$ (ETO) (Fig. 1a) [11, 12] and $LaMnO_3$ [13]. FM correlations were reported also in heterostructures where LAO is replaced by a ferro(ferri)-magnetic insulator, as in $GdTiO_3$/STO (13, 14), EuO/STO [16] and $LaAl_{1-x}Mn_xO_3$/STO [17].

In LAO/ETO/STO, FM correlations are believed to be induced by the ordering of localized $Eu^{2+}$ $4f^7$ magnetic moments and their coupling with Ti-$3d$ states forming the conduction band of the q2DEG. However, the microscopic mechanism leading to these phenomena is not straightforward, as rare earth magnetic ions have usually a small hybridization with transition metal $3d$ orbitals. Unlike the LAO/STO q2DEG [1-9], there is no detailed study of the electronic band structure of LAO/ETO/STO heterostructures, and it remains unclear whether a spin-polarized q2DEG is present in both ETO and interfacial STO layers, which is relevant for the intriguing phase diagram showing a transition from a FM to a superconducting state [12]. Additionally, FM correlations were observed only above the Lifshitz transition, where carriers with $3d_{xz,yz}$ orbital character start to contribute to the transport [12].

In this work all these questions are settled by combining extensive experimental and theoretical investigations. Firstly, we show that low temperature electrical transport, x-ray magnetic circular dichroism (XMCD) and superconducting quantum interference device (SQUID) experiments give evidence of tunable ferromagnetism originating from the correlation between Ti- and Eu- magnetic moments. Secondly, we provide a direct picture of the q2DEG band structure by using resonant soft-x-ray photoemission spectroscopy (RESPES). These experimental observations are finally compared and blended with Density Functional Theory calculations with on-site Hubbard terms (DFT+$U$). The latter show that the q2DEG forms at the ideal LAO/ETO interface and extends few unit cells into the STO. FM correlations are induced by the delicate balance between different antiferromagnetic (AFM) and FM contributions to the exchange interaction between Eu-$4f^7$ magnetic moments and Ti-$3d$ electrons at the ETO/STO interface, which involves also the first STO unit cells (uc), in agreement with the experiments. Moreover, theoretical calculations show that the ferromagnetism in the q2DEG and the filling of $3d_{xz,yz}$ bands take place simultaneously, explaining the electric field induced anomalous Hall effect above the Lifshitz point [12], and the orbital selective switching of the ferromagnetism in an oxide q2DEG.

**Results**

**I. Experimental evidence of intrinsic ferromagnetic correlations**

Bulk ETO is an AFM insulator with a Neel temperature of 5.5 K. However, doping or lattice strain favor a FM order in epitaxial films [18-22]. ETO has a quasi-perfect structural matching with STO and a similar conduction band formed by the overlap of $t_{2g}$ Ti-$3d$ orbitals. By embedding few unit cells of ETO between LAO and STO using epitaxial growth by pulsed laser deposition, we realized a q2DEG characterized by electrical transport properties similar to the LAO/STO q2DEG. However, at low temperatures and at a carrier density $>2.10^{13}$ cm$^{-2}$ tuned by electric field effect (Fig.1a), the sheet resistance exhibits a downturn below T=6-8 K (Fig.1b), which correlates to the FM transition of doped ETO [12]. The heterostructures itself is characterized by a FM transition ($T_c$=6-8 K) and a FM ground state, as shown in Fig.1c, where we compare the maximum intensity of the Eu-XMCD signal (at $M_5$-edge, scatter data), acquired at 2 K at the X-Treme beamline of the Swiss Light Source (SLS) [23], to macroscopic SQUID magnetometry data (full lines) as a function of the magnetic field in parallel (red line) and perpendicular (black line) directions. The two sets of data, normalized to the saturation value for direct comparison, show a small, although clear, hysteresis around zero magnetic field and a preferential orientation of the magnetization parallel to the interface, confirming a finite low magnetic field magnetization and FM correlations. The magnetization saturates above 2-3 Tesla (depending on the sample and on the magnetic field direction), indicating that all the Eu$^{2+}$-spin moments order ferromagnetically above these values.

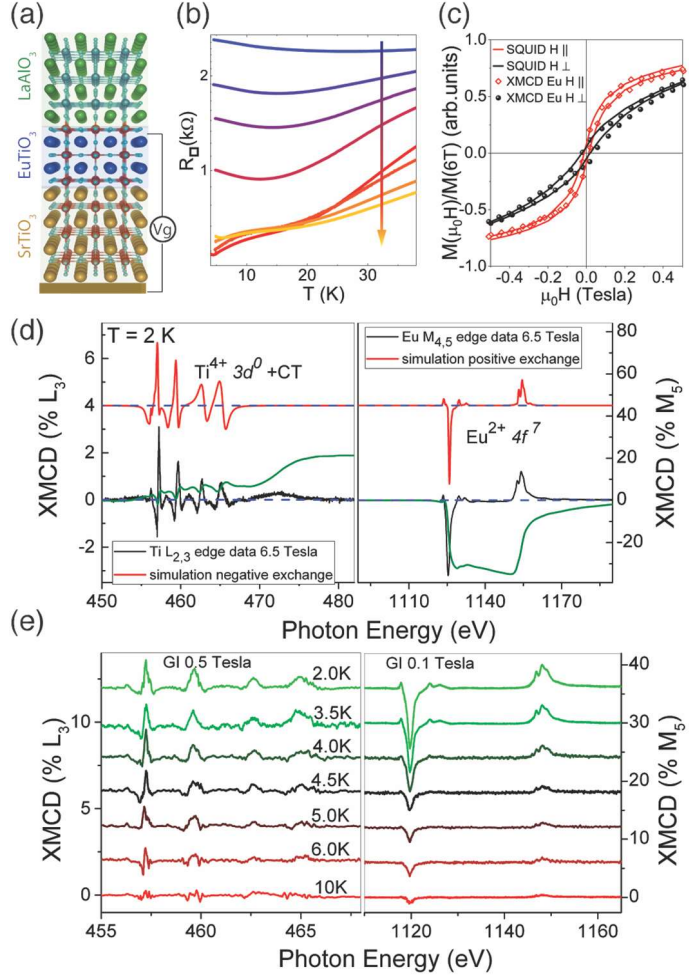

**Fig. 1. Structure and general transport and magnetic properties of LAO/ETO/STO heterostructures. (**a) Sketch of the LAO/ETO/STO heterostructure. In this work, the carrier concentration was tuned using electric field effect in back-gate configuration, as sketched in the figure. (b) Sheet resistance vs. temperature as a function of the gate voltage Vg. The arrow indicates increasing values of Vg from -30 to 30 V. (c) Eu-XMCD (scatter data) and SQUID magnetization (continuous lines) as function of the magnetic field parallel (red) and perpendicular (black) to the interface. The data are normalized to the saturation value. The Eu-XMCD in parallel direction is obtained by combing data acquired in grazing (60 degrees from the surface normal) and normal incidence conditions. (d) XMCD data at the Ti-L (right) and Eu-M (left) edges at 6.5 Tesla. Red-lines are atomic multiplet simulations, green lines are integrals of the XMCD spectra (see main text). (e) Temperature dependence of the Ti-$L_{2,3}$ (0.5 Tesla) and Eu-$M_{4,5}$ (0.1 Tesla) edges XMCD spectra in grazing incidence conditions. Data are vertically shifted for clarity.

To further elucidate the magnetic properties of the system, in Fig.1d we show XAS and XMCD spectra at the Ti-$L_{2,3}$ and Eu-$M_{4,5}$ edges at 2 K and 6.5 Tesla and in Fig.1e the XMCD temperature dependence in grazing incidence conditions (60 degrees from the surface normal). The data show that Eu- and Ti- XMCD spectra follow each other, thus are correlated. In particular the Ti- (orbital and spin) and the Eu- magnetic moments have the same temperature dependence, with a $T_c$ of the order of 6-8 K. Moreover, they have the same magnetic field dependence as shown in ref. [12].

In Figure 1d, the XMCD spectra are compared to atomic multiplet simulations (red lines) for $Eu^{2+}$ and $Ti^{4+}$ ions in C4 symmetry, the latter including the crystal field splitting reproducing the x-ray linear dichroism data, and a charge transfer (CT) term to account for the hybridization between Ti-*3d* and O-*2p* states in the $TiO_6$ cluster (see Supplementary materials from page 24, and ref. [24]). The calculations reproduce most of the features shown by the experiment. However, while a positive exchange is needed to simulate the $Eu^{2+}$ magnetization, parallel to the magnetic field, only a negative one correctly reproduces the Ti-XMCD peaks at $L_3$ and $L_2$. On the other hand, we also find a finite and positive Ti-XMCD integral that, according to the sum rules [25], corresponds to a sizeable Ti-*3d* orbital moment, $m_{orb}$, of the order of -0.05 $\mu_B$/Ti, opposite to the magnetic field. While the application of the sum rules does not accurately provide the spin-moment at the Ti-$L_{2,3}$ edge, following the Hund's rules, the sign of the orbital moment corresponds to a finite Ti-spin moment parallel to the Eu-spin moment.

However, the Ti-XMCD integral and the Ti- magnetic moment is expected to be null for a Ti-*3d⁰* state, even in the presence of a CT between Ti-*3d* and neighbor O-*2p* states. Thus, the only way to explain the results is the presence of two contributions to the Ti-XMCD spectra: (i) a *3d⁰*+CT contribution (i.e. *3d⁰* and *3d¹L*, with *L* the O2p-Ti3d ligand state), due to the negative exchange interaction between the ligand Ti*3d* electrons and the $Eu^{2+}$ magnetic moments, which gives rise to a circular polarization dependence of the cross-section for the *2p⁶3d⁰* (*3d¹L*) to *2p⁵3d¹*(*3d²L*) transition; (ii) a second contribution coming from Ti-*3d¹* electronic states, with their associated magnetic moments, parallel to the magnetic field and to the $Eu^{2+}$ spin-moments. These results suggest different typologies of exchange interactions between Eu and Ti ions i.e., an AFM one between $Eu^{2+}$ and Ti-*3d⁰*+CT states, and a FM exchange between the $Eu^{2+}$ and Ti-*3d¹* spin-moments.

The FM correlations in the q2DEG, and the magnetic field dependence of the magnetization are directly related to the Hall effect data [12], which exhibit at low field (<3 Tesla) a curvature change associated to the anomalous Hall effect (AHE) (Fig.2a). Further insights on the nature of the AHE is obtained by comparing the 2D anomalous conductivity, $\sigma_{AHE}$, to the longitudinal one, $\sigma_{xx}$, estimated according to eq.(*1a,1b*):

$$\sigma_{xx} = \frac{\rho_{xx}^{2D}(0)}{(\rho_{AHE})^2 + \left(\rho_{xx}^{2D}(0)\right)^2} \quad (1a)$$

$$\sigma_{AHE} = \frac{\rho_{AHE}}{(\rho_{AHE})^2 + \left(\rho_{xx}^{2D}(0)\right)^2} \quad (1b)$$

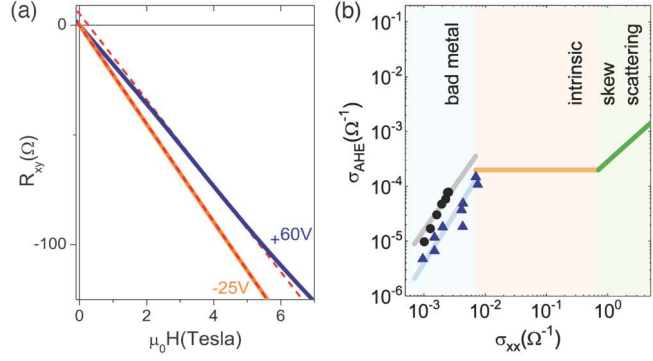

**Fig. 2. Analysis of the anomalous Hall effect.** (a) Transverse (Hall) resistance $R_{xy}$ as a function of the magnetic field for two values of the gate voltage. The high gate voltage +60 V (higher carrier concentration) data show negative curvature at low field ($\mu_0H<2.5$ T) and a positive curvature at high field ($\mu_0H >5T$). The former is due to the anomalous Hall component, the latter to multiband transport. The red dashed lines are linear fit in the range 3-4 Tesla, and the intercept is the AHE component at saturation. (b) Anomalous Hall conductivity vs. longitudinal 2D conductivity (see main text eq.1). Black dots and blue triangles refer to two LAO/ETO/STO samples. Full lines refer to the exponent $\alpha$ in the relation $\sigma_{AHE} \propto (\sigma_{xx})^\alpha$: $\alpha$=1.8 (light blue, and gray, intrinsic AHE suppressed by disorder), $\alpha$=0 (orange, intrinsic AHE), $\alpha$=1 (green, skew scattering).

where $\rho_{xx}^{2D}(0)$ is the zero field 2D resistivity, and $\rho_{AHE}$ is the saturation value of the anomalous Hall 2D resistivity, as calculated from the anomalous component of the Hall effect (Fig. 2a and refs. [12, 19]). The origin of the AHE can be attributed to different mechanisms, which can be phenomenologically distinguished from the relationship between $\sigma_{AHE}$ and $\sigma_{xx}$. According to the literature [26], most of the materials showing an AHE can be classified in the following three regimes: (i) a high conductivity range, where $\sigma_{AHE} \propto \sigma_{xx}$ and the observed AHE is attributed to scattering from unwanted magnetic impurities (skew scattering); (ii) an intermediate conductivity range, where $\sigma_{AHE}$ is independent from $\sigma_{xx}$ and the AHE is related only to the Berry curvature of the involved bands (intrinsic AHE); (iii) a bad metal range where $\sigma_{AHE} \propto (\sigma_{xx})^\alpha$, with $\alpha$ in the interval 1.6-1.8, where the intrinsic AHE is suppressed by the disorder. In the latter two ranges the AHE is due to the spin-polarization of the carriers and to the spin-orbit coupling breaking both time and inversion symmetry. Data collected on two LAO/ETO/STO representative samples, with carrier densities tuned by a back-gate voltage, are shown in Fig.2b. We found a correlation $\sigma_{AHE} \propto (\sigma_{xx})^\alpha$, with $\alpha$=1.8, which excludes a skew scattering mechanism and a purely intrinsic AHE. We conclude that in LAO/ETO/STO an intrinsic AHE suppressed by disorder is at play, analogously to what found in La-doped ETO films [20].

**II. Experimental investigation of the band structure via RESPES**

In order to characterize the electronic structure of the LAO/ETO/STO system, we used the RESPES technique at the soft-x-ray end station of the ADvanced RESonant Spectroscopies (ADRESS, X03MA) beamline of the Swiss Light Source (SLS) [27, 28]. The experimental data were acquired at base temperature (12 K) thus above the FM transition of the system.

In Fig. 3a we report angle and photon-energy integrated (across the Ti-$L_{2,3}$ edge) RESPES VB data on (001) LAO(5uc)/ETO(2uc)/STO and a reference PLD grown (001) LAO(5uc)/STO sample containing oxygen vacancies, i.e. cooled down from the high deposition temperature (750 °C) in a reduced oxygen pressure of the order of $10^{-5}$ mbar, without any high-$O_2$ pressure annealing process. For comparison we show also data acquired by high resolution ARPES at the SIS (Surface Interface Spectroscopy) beamline of SLS ($h\nu$ = 85 eV), on reference VB of an STO crystal and of an ETO film hosting q2DEGs [35].

Since the escape depth of the photoelectrons in the 450-470 eV range is of the order of 1-2 nm, in LAO(5)/ETO(2)/STO the data have contributions from both the ETO layers and from the interfacial STO unit cells. As a result, the VB spectra of ETO heterostructures show prominent differences respect to the STO and LAO/STO VB. The main one is the presence, in ETO and in LAO/ETO/STO, of a peak at -1.95 eV binding energy (BE), due to $Eu^{2+}$ - $4f^7$ states. Another important difference between the STO and LAO/STO q2DEGs containing oxygen vacancies, and the ETO q2DEGs is the absence in the latter of the -1.0 eV in-gap state (IGS). The IGS is a distinctive characteristic of the STO q2DEGs and of oxygen deficient LAO/STO [28]. Finally, in LAO/ETO/STO we observe the evidence of a further broad feature at BE lower than -8 eV, which reflects the contribution of Eu in $Eu^{3+}$ oxidation state: through a Gaussian deconvolution of the observed spectral features (reported in Fig. 3a), we estimated a fraction of $Eu^{3+}$ of about 25%. This result is in agreement with reported atomically resolved high resolution transmission electron microscopy and electron energy loss spectroscopy [12], which show that a similar fraction of $Eu^{3+}$ is present in the system, mostly located into the LAO layer at the interface with the ETO film.

In Fig. 3b we report angle integrated VB-RESPES data of LAO(5uc)/ETO(2uc)/STO at photon energies $h\nu$ across the Ti-$2p$ absorption edge resonance. The data are overlapped to TEY XAS (white line), and to Constant Initial State (CIS) spectra obtained by integrating the valence band over different binding energy ranges. CIS spectra allows the identification of the resonance of different VB features with the XAS intensity spectra. We can see that the same XAS resonances characteristic of $Ti^{4+}$ ions appear as higher intensity signal in the RESPES map inside the contribution from the O-*2p* band. Similar results have been reported for the LAO/STO q2DEG [28], and are a consequence of the hybridization between O-*2p* and Ti-*3d* states. On the other hand, the CIS spectrum around the Fermi level, related to the q2DEG conduction band, has a different shape, reminding the characteristic absorption from $Ti^{3+}$ ions in an octahedral environment, similar to bulk $LaTiO_3$ [29] (see Supplementary Materials, from page 24). Surprisingly, also the peak at -2 eV resonates with the Ti-absorption spectrum, a result which would point to a hybridization of Eu with O-*2p* and Ti-*3d* states. This is rather unexpected, since *4f* rare earth orbitals have usually little overlap with neighbor ions in a crystal.

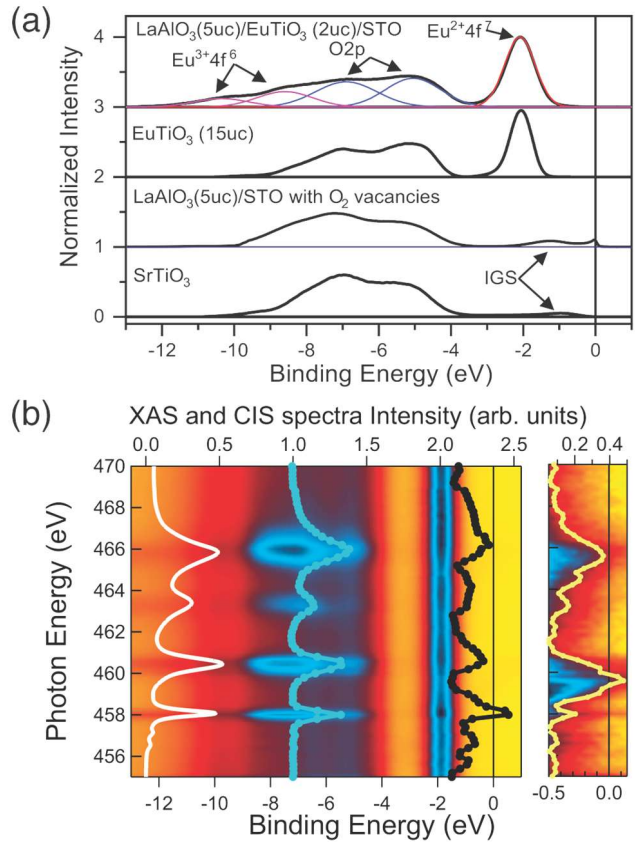

**Fig. 3. RESPES and ARPES VB data**. (a) Angle integrated ARPES VB data of $SrTiO_3$, and $EuTiO_3$ q2DEG surface state, compared to angle and photon-energy integrated RESPES VB of (001) LAO(5uc)/STO (containing oxygen vacancies) and LAO(5uc)/ETO(2uc)/STO heterostructures. Multiple gaussian fit of the LAO/ETO/STO VB profile is used to determine the $Eu^{3+}/Eu^{2+}$ ratio. The IGS in gap states is observed only in STO and LAO/STO q2DEGs due to the presence of oxygen vacancies. (b) (left panel) RESPES color-map of the valence band region and constant initial state spectra obtained integrating the RESPES data around BE = -2 eV ($Eu^{2+}$, black scatter data) and between -7 and -4 eV (O-*2p*, cyan scatter data). The white line is the total electron yield XAS spectrum; (right panel) the same RESPES map in the Fermi level region and corresponding CIS spectrum (yellow scatter data).

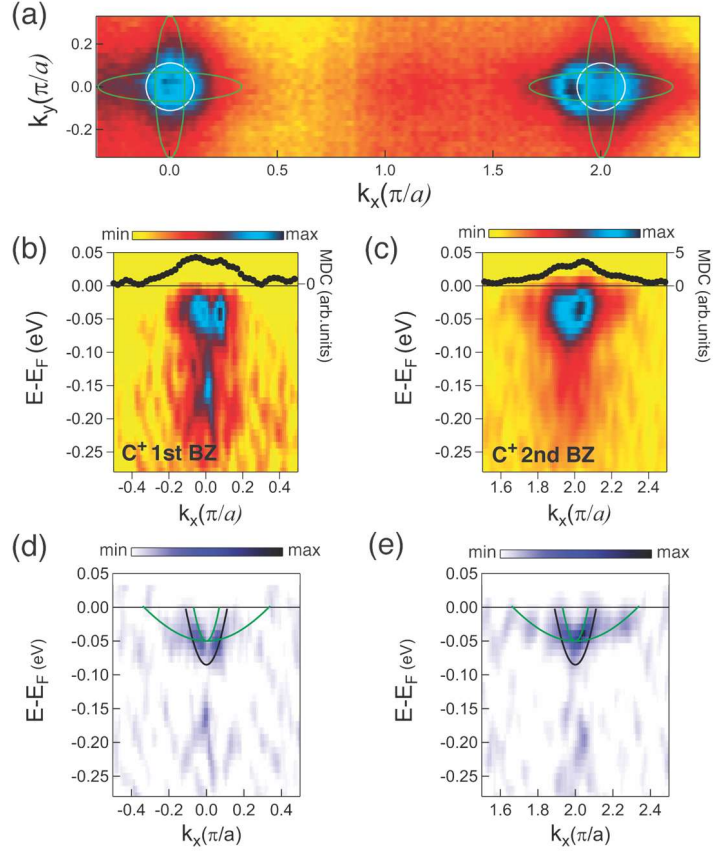

**Fig. 4: Resonant angle resolved photoemission spectra.** (a) Average between circular plus and circular minus Fermi surface map of the (001) LAO/ETO/STO heterostructure. Dashed lines are Fermi surface contours of light (white) and heavy (green) bands. (b) and (c) E vs. $k_x$ band dispersions maps acquired with C+ polarization in the 1st and 2nd BZ. Black scatter data on top of each band dispersion map are the Fermi level MDCs curves (integrated around $E_F$ in a 10 meV range). All the data are measured in resonance conditions (465.5 eV). (d) and (e) two-dimensional curvature maps of the C+ dispersions in (c) and (d), with tight-binding fitting of the bands (green lines $3d_{xz}$ and $3d_{yz}$ bands, black line $3d_{xy}$ band).

In order to characterize the band structure of the LAO/ETO/STO q2DEG, in Fig. 4a we show $k_x$-$k_y$ in-plane cut of the Fermi Surface (FS), obtained by averaging circular minus (C-) and circular plus (C+) polarization spectra at an incoming photon energy of 465.5 eV, resonant with the $Ti^{3+}$ $L_2$ absorption peak. Overlapped to the map we also show Fermi surface contours associated to non-interacting electronic bands originated from atomic Ti-$3d$-$t_{2g}$ states: namely, a ring-shaped feature and two ellipsoidal structures oriented lengthwise along the $k_x$ and $k_y$ directions. The ring-shaped Fermi contour corresponds to light effective mass electrons having mainly $3d_{xy}$ orbital character, while the ellipsoidal ones are related to heavy effective masses electrons with mainly $3d_{xz}$-$3d_{yz}$ orbital characters. The qualitative features of the measured FS in (001) LAO/ETO/STO resemble those observed on different oxide systems characterized by the presence of a q2DEG, like the (001) STO surface [30, 31], the (001) LAO/STO interface [32-34], and the recently investigated (001) ETO surface [35]. However, the FS, in particular at the 2nd BZ, where heavy bands contributions are better resolved due to matrix element effects, has a shape which departs from the simple tight binding model of three non-interacting $t_{2g}$ bands.

In Figs. 4b-c we show band dispersion cuts at the 1st and 2nd BZ through the Γ-point along the $k_x$ direction (corresponding to Γ-X in reciprocal space) with circular plus (C+) polarization of the incoming photons (C+ pol, hν = 465.5 eV). Momentum Dispersion Cuts (MDCs) at the Fermi level are shown on the top of each panel. In Fig. 4d-e we show the corresponding 2D-curvature maps to highlight the different bands [36]. Other data acquired with different polarizations of the incoming photons are shown in the Supplementary Materials (from page 24).

The dispersive profile of the different bands were obtained by combining the 2D-curvature maps and the fit of the maxima of MDCs at several energies below the Fermi level, in analogy with previous studies on LAO/STO q2DEGs [33, 34] (see Supplementary Materials, from page 24). From simple tight-binding fitting assuming three independent bands (superimposed to the 2D curvature maps of Fig. 4d-e), we estimated the effective masses m* at the Fermi level of each band (see Table 1). We obtained m* = 0.4 $m_e$ for the $3d_{xy}$ band, while for the heavy bands we estimated m* = 0.25 $m_e$ and ~10 $m_e$ in the "light" and "heavy" directions, respectively ($m_e$ being the free electron mass). The $3d_{xy}$ band effective mass and the $3d_{xz,yz}$ bands effective masses along the "light" directions are typically smaller than the ones estimated for the LAO/STO system [33]. The splitting between the heavy- and light- band bottoms is of the order of ~ 35 meV at the Γ-point, considerably smaller than that usually reported for the STO-surface 2DEGs [30], but consistent with our earlier reports about the differences between the ETO and STO surface states [35], and in quantitative agreement with x-ray linear dichroism data on (001) LAO/ETO/STO [12].

Table 1: Fermi momentum $k_F$, band bottom E(0) and effective mass of the various bands.

| Bands | $k_F$ (Å$^{-1}$) | E(0) (meV) | m*/$m_e$ |
|---|---|---|---|
| $3d_{xy}$ | 0.09±0.01 | -85±5 | 0.4±0.1 |
| $3d_{yz}$ | 0.27±0.02 | -50±5 | 10±2 |
| $3d_{xz}$ | 0.055±0.005 | -50±5 | 0.25±0.05 |

## III. Results of DFT+*U* calculations

In order to understand the mechanism at the base of the formation of the q2DEG in LAO/ETO/STO and of the spin-polarization of its carriers, we performed DFT+*U* calculations [37-42] on an ideal c(2x2) LAO/ETO/STO (001) heterostructure composed by 5uc of LAO, 2uc of ETO and 4uc of STO stacked along the c-axis, and a vacuum region of 20 Å (Fig. 5a). On-site effective Hubbard parameter $U$ = 4 eV, 7.5 eV and 8 eV were applied on the Ti-*3d*, Eu-*4f* and *La-4f* states, respectively. Similar to findings for the ETO (001) surface state [35], the choice of $U$ for the Eu-*4f* states, and partially for the Ti-*3d* states, is dictated by the necessity to reproduce the position of the $Eu^{2+}$ peak in the valence band at about -2 eV. We tested models with the Eu-ions in AFM (G-type) and FM configurations. We found that the FM solution is the lowest in energy for the system (about -20 meV per simulation cell). This result is in full agreement with the experimental evidence of a FM ordering of $Eu^{2+}$ magnetic moments shown in Figs.1-2 and in ref.[12].

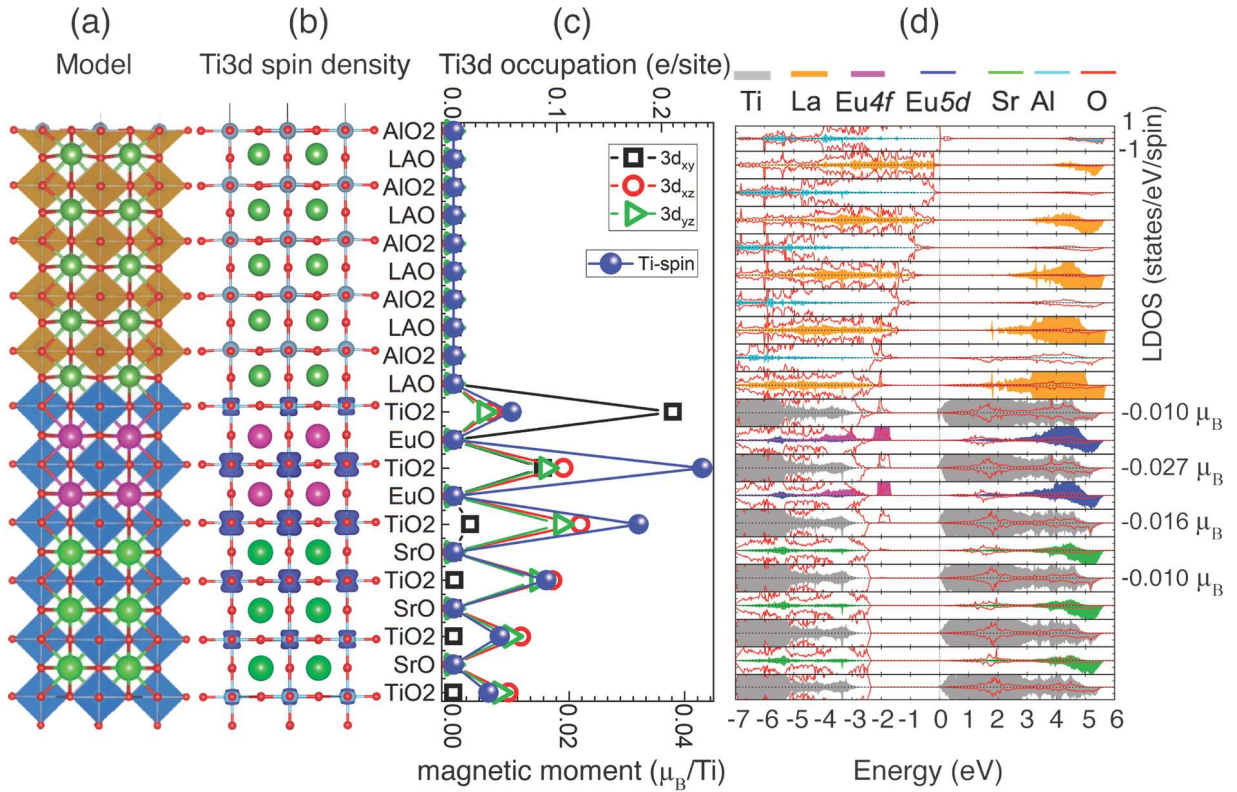

**Fig. 5. DFT+U calculations:** (a) Structural model of a c(2x2) LAO/ETO/STO(001) heterostructure; (b) integrated Ti-*3d* spin density plot with isovalue of 0.0004 e/Å$^3$ across the interface. (c) orbital and layer resolved Ti-*3d* occupation (upper scale; black squares, $3d_{xy}$; red circles, $3d_{xz}$, green triangles, $3d_{yz}$) and Ti-*3d* magnetic moment (lower scale; blue circles) obtained by integrating the density of states between -0.3 eV and $E_F$. d) Layer-, spin- and element- resolved density of states. The color code for the partial LDOS contribution of different ions is indicated on top of the figure. On the right side of the figure, we also indicate the values of the calculated oxygen-*2p* magnetic moment in the corresponding interfacial $TiO_2$ layers.

In Fig. 5 we show the Ti-*3d* spin density (Fig. 5b), the orbitally resolved Ti-*3d* occupation and magnetic moments (Fig. 5c), and the layer, atomic and spin resolved, local density of states (LDOS, Fig. 5d) across the interface. According to the DFT+*U* results, the q2DEG at the LAO/ETO/STO interface is formed through the transfer of electrons from the $AlO_2$ surface layer of LAO to the Ti-states in the ETO and STO layers in order to eliminate the polar discontinuity at the LAO/ETO interface. This is evidenced by the overlap between the energy positions of the O-*2p* band of the $AlO_2$ surface and of the interfacial $TiO_2$ *3d*-conduction bands in Fig. 5d, analogously to what happens in LAO/STO bilayers. We underline that ETO planes are formally neutral along the (001) direction alike STO, as Eu is mostly in $Eu^{2+}$ valence state. Figure 5c shows that the q2DEG is characterized by a high occupation of Ti-*3d* states within the ETO layer, which goes to zero within few unit cells of STO. $3d_{xy}$ bands are the first to be occupied at the interface and are mostly localized into the ETO film.

The bands are spin-polarized, with the highest polarization within the ETO layers, which exhibit also the largest electron occupation. In particular, the model shows that most of the electrons filling the STO unit cells have a $3d_{xz,yz}$ orbital character, reflecting the wider distribution of these carriers, as also observed in (001) STO [30]. The Ti-*3d* magnetic moment, obtained by integrating the spin-resolved density of states in the range between -0.3eV and $E_F$ (i.e. the contribution from spin-polarized conduction bands) is maximum in the second ETO layer, and then goes to zero within the first three-unit cells of STO (Figure 5c). Consequently, both ETO and topmost STO unit cells host electrons which are spin-polarized. The associated magnetic moment is parallel to the large spin-moment of $Eu^{2+}$ states (6.97 $\mu_B$). The layer resolved map in Figure 5d shows also a finite spin-polarized electron density of Ti-*3d,* Eu-*5d* and O-*2p* states at the position of the $Eu^{2+}$ peak, demonstrating a hybridization between Eu-*4f*, Eu-*5d*, O-*2p* and Ti-*3d* states, in agreement with RESPES data. Moreover, O-*2p* states in the $TiO_2$ layers of ETO also acquire a finite spin-moment, opposite to the Ti-*3d$^1$* and Eu-*4f$^7$* magnetic moments. This can explain the two components of the XMCD spectra, one related to the Ti-*3d$^0$* +CT states, with a negative exchange interaction with the $Eu^{2+}$ *4f* states, and the other from Ti-*3d$^1$* spin-polarized electrons, parallel to the overall magnetization direction.

To shed light on the bands contributing to the q2DEG, we show in Fig. 6a the calculated, spin resolved, band structure of LAO/ETO/STO (001) in the FM ground state in a large energy range around $E_F$, including the $Eu^{2+}$-*4f* state around -2 eV. Theory predicts several non-spin-polarized bands in the -1.5 eV to $E_F$ binding energy range deriving from the LAO contribution to the density of states, including a (hole-) band (green line in the figure) crossing the Fermi level associated with the $AlO_2$ surface state (LAO SS). The LAO related bands are not observed in the RESPES data, in agreement with previous studies on standard LAO/STO interfaces [32-34], indicating that the LAO exposed surface is not ideal and can be passivated by adsorbates, in particular after ambient exposure.

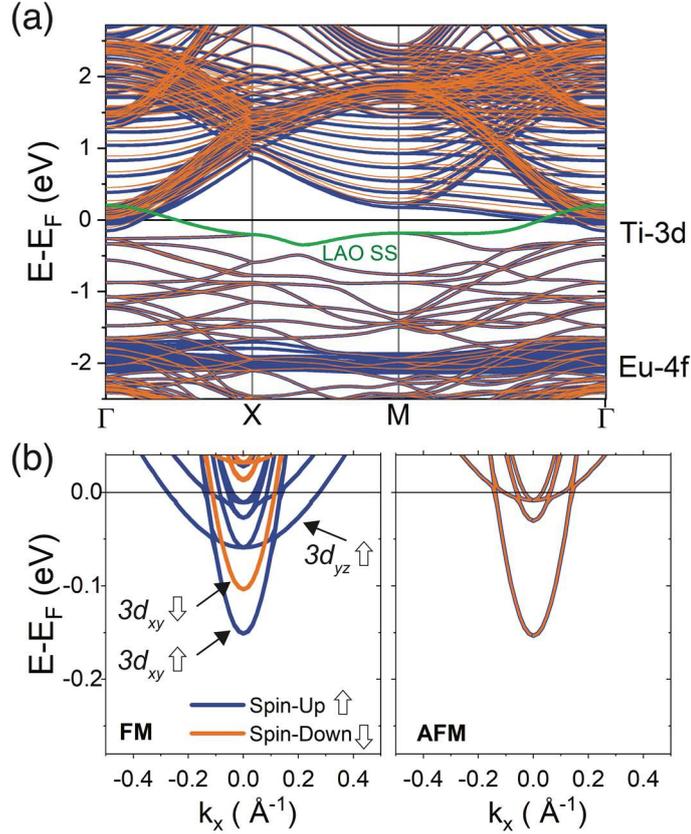

**Fig. 6. Spin-resolved band structure from the DFT+$U$ calculations**: (a) Band structure DFT+$U$ calculations on LAO/ETO/STO system in the FM ground state (lowest energy configuration). Spin-up/spin-down bands are denoted in blue/orange. The LAO surface state (LAO SS) is outlined as a green line. The Eu-*4f* band around -2 eV is fully spin-polarized. In the figure Γ-X-M refer to the c(2x2) unit-cell notation (Note that Γ–M in the c(2x2) unit cell is equivalent to Γ–X in the (1x1) unit cell). (b) Spin-resolved band structure calculations of the Ti-*3d* bands around the Γ-point in a small energy range in the FM (left panel) and AFM (right panel) solutions of the model. Note that in the FM solution, the spin-up *3d$_{yz}$* band shift to lower energy and crosses the Fermi level, while the *3d$_{xy}$* spin down bands shifts-up, giving rise to an overall spin-polarization of the q2DEG.

In order to compare the experimentally determined band structure shown in Fig. 4 with DFT+$U$ calculations, and to highlight the differences between the conduction bands in the FM (the ground state) and AFM solutions of the model, we show in Fig. 6b the calculated Ti-*3d* band structure in the two cases (left panel FM, right panel AFM) in a small energy region around the Fermi level. In the case of the AFM state, we have forced a (G-type) ordering of the Eu-spin moments. It turns out that the calculated band structures are qualitatively consistent with the main characteristics of the experimentally determined band properties shown in Fig. 4. In both the FM (Fig.6b, left panel) and AFM (Fig.6b, right panel) solutions, the lowest lying Ti-*3d* band crossing $E_F$ is a dispersive parabolic band with $d_{xy}$ orbital character, which switches at larger momentum into a much flatter band indicating an avoided crossing with the heavy bands with $d_{xz}$, $d_{yz}$ orbital character. The higher-lying bands around $\Gamma$ are replicas from different layers due to the confinement of the q2DEG. In the FM solution of the model, spin-up (blue) and spin-down (orange) bands, both of the $d_{xy}$, and $d_{xz,yz}$ main orbital character, are split in the whole energy range and cross the Fermi level, showing a prevalence of majority spin-polarized electrons in the system, in full agreement with all the experimental evidences of FM correlations in the q2DEG. In the AFM case the spin-degenerate $3d_{xy}$ bands arrange in the same energy region as the spin-up band of the FM phase, while $3d_{xz,yz}$ bands are lifted above the Fermi level. Thus, the conduction band minimum, associated with the lowest $3d_{xy}$ band is the same in both AFM and FM phases, but all the other bands are shifted in the FM solution. DFT+$U$ calculations in particular show a down-shift of the spin-up $3d_{xz,yz}$ bands and more importantly a filling of these bands taking place only in the FM ground state. Thus, ferromagnetism appears simultaneously with the orbital selective filling of $3d_{xz,yz}$ electrons at the Fermi level, confirming their crucial role in establishing FM correlations in the q2DEG as earlier suggested in ref. [12].

**Discussion**

The experimental and theoretical results on LAO/ETO/STO (001) heterostructure show an interaction between $Eu^{2+}$-*4f* and Ti-*3d* states. This is rather surprising considering the large value of the Hubbard parameter on the $4f^7$ orbitals, and their strong electron localization, rendering them not far from a configuration with electrons frozen in the core.

In bulk $EuTiO_3$, the establishment of a FM ground state, instead of an AFM one, is believed to be due to a delicate balance between the different exchange interactions among $Eu^{2+}$ magnetic moments in the system: i) Direct FM exchange between $Eu^{2+}$, which is very weak due to the almost null overlap between Eu-*4f* states; ii) an AFM super-exchange interaction mediated by O-*2p* states: iii) an AFM super-exchange interaction mediated by Ti-*3d* states; iv) a carriers mediated Ruderman-Kittel-Kasuya-Yoshida (RKKY)-like indirect FM-exchange, proposed for La-doped FM ETO films [18], due to an overlap between (filled) $t_{2g}$ Ti-*3d* states and Eu *4f*-orbitals; and v) an indirect exchange interaction via Eu-*5d* $t_{2g}$ states, which play an important role in Eu-chalcogenides (e.g.

EuO). It has been proposed that the filling of the Ti-*3d* bands reverts the indirect exchange via the Eu-*5d* $t_{2g}$ states from AFM to FM through the coupling between $t_{2g}$ Eu-*5d* and Ti-*3d* electrons [43, 44]. Our experimental XMCD data confirm that the super-exchange between $Eu^{2+}$ magnetic moments through O-*2p* and Ti-*3d* states is negative, thus the only two mechanisms which could effectively give rise to ferromagnetism are FM-couplings mediated by itinerant carriers (without any role of Eu-*5d* states) or mediated by Eu-*5d* hybridized with Ti-*3d* states [44].

In order to verify if the latter can effectively play a role in our heterostructures, in Fig. 7a we show the average spin-resolved LDOS of the ETO layers in the FM ground state. The calculations are compared to the LDOS in the AFM configuration (Fig.7b). We find that Eu-*5d* states, while in general characterized by a much weaker spectral weight than the O-*2p* and Ti-*3d* contributions, have a strong overlap with Ti-*3d* $t_{2g}$ bands in the FM ground state close and above the Fermi level. Furthermore, there is also a substantial overlap with the Eu-*4f* state around -2 eV, where these states also show a clear spin-polarization. On the other hand, in the AFM solution the Eu-*5d* states are at higher energy, have a lower spectral weight compared to the FM-case, and their overlap with Ti-*3d* states is substantially reduced. These results suggest that the FM-ordering of $Eu^{2+}$ is mostly due to a FM interaction mediated by Eu-*5d*/Ti-*3d* hybridized states as suggested earlier [44], although it is not possible to fully exclude a role of a carrier mediated RKKY- FM interaction [18].

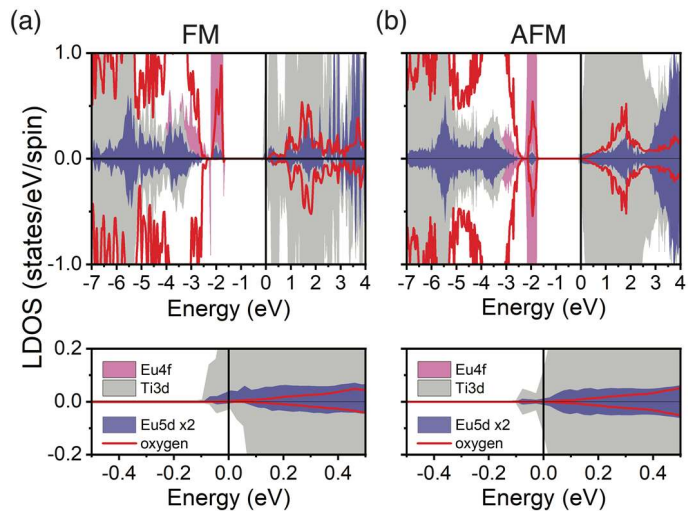

**Fig. 7. Comparison between AFM and FM solutions of the model:** (a)-(b) Element and spin-resolved density of states from the DFT+*U* calculations for $EuTiO_3$ layers (average) with (a) FM and (b) AFM coupling. In the FM solution we do observe a substantial spin-polarization at -2 eV of both Eu-*4f* (magenta) and Eu-*5d* (blue) states, and an overlap between Ti-*3d* states (grey) and Eu-*5d* states near and above the Fermi level, absent in the AFM case.

Very interestingly the calculations also demonstrate that the filling of Ti-$3d_{xz,yz}$ bands and FM-order appear at the same time, as shown in Fig.6b. Therefore, Ti-$3d_{xz,yz}$ electrons have the main role in the establishing a FM-coupling among Eu$^{2+}$ magnetic moments in the confined heterostructure, while $3d_{xy}$ electrons seems less effective in mediating a dominant FM interaction in this system.

**Conclusions**

In conclusion, in this work we combined different experimental methods and theoretical calculations to analyze and clarify the origin and properties of the spin-polarized q2DEG at the LAO/ETO/STO oxide interface. The DFT+$U$ calculations show that a FM and spin-polarized q2DEG is formed at the (001) LAO/ETO/STO heterostructure in a defect-free interfaces. At the same time, theory explains some crucial properties of this q2DEG revealed by x-ray spectroscopy and electrical transport results. In particular, it is shown why the filling of $3d_{xz,yz}$ bands, inferred from the experimental results, is simultaneous to the transition to a FM state, thus explaining why the q2DEG is spin-polarized only above the Lifshitz transition. Moreover, spin-polarized $3d_{xz,yz}$ electrons created at the LAO/ETO interface leak also into the first layers of STO, explaining the contribution from STO to the Ti-$3d$ magnetic moment found in previous investigations [12].

Our results show that the LAO/ETO/STO system provides a platform for the study of novel quantum phenomena where superconductivity, magnetism and spin-orbit coupling are fully entangled, and is a clear example on how new functional properties can be created in oxide 2D-systems by atomic interface engineering.

More generally, the combination of electrical transport and spectroscopy measurements with band structure calculations reported in this work is a powerful tool to obtain a deeper understanding of complex heterostructures characterized by unexpected novel functional properties. Such a combined approach is becoming essential to gain detailed understanding of novel heterostructures showing exceptional properties.

**Materials and Methods**

LAO(n)/ETO(2)/STO(001) heterostructures were fabricated by pulsed laser deposition (PLD) assisted by Reflection High Energy Electron Diffraction (RHEED) from sintered Eu$_2$Ti$_2$O$_7$ and crystalline LAO targets onto TiO$_2$-terminated (001) STO substrates. The samples were deposited at a temperature of 700 °C in a background O$_2$ pressure of 8 × 10$^{-5}$ mbar and cooled down in the same conditions to room temperature with a rate of 5°C/min. We used an excimer laser (Lambda Physics, 248 nm wavelength) and 1.3 J/cm$^2$ fluence, 1Hz of repetition, resulting in a rate of 0.05 unit cell/pulse (20 pulses for each unit cells). A q2DEG is formed when n> 4 unit-cells (uc).

We have used polarization dependent x-ray absorption spectroscopy (XAS) across the Eu $M_{4,5}$ and the Ti-$L_{2,3}$ edge to probe directly the magnetic and orbital properties of Eu and Ti at the interface. The experiments were performed at the beamline X-Treme of the Swiss Light Source [23]. XAS performed with circularly or linearly polarized photons can detect the magnetic moments and the *3d* orbital energy splitting, respectively. The two techniques, usually known as x-ray magnetic circular dichroism (XMCD) and x-ray linear dichroism (XLD) are so sensitive that they can be used on single interfaces. The Eu $M_{4,5}$ edge and the Ti $L_{2,3}$ XMCD spectra were obtained as difference between the average of 8 and 16 (respectively) XAS spectra acquired with magnetic field parallel and antiparallel to the photon-helicity vector orientations. The 16 and 32 XAS data needed for each XMCD were collected in a sequence alternating reversal of field and polarization at each spectrum. This procedure ensures the best cancellation of spurious effects. The magnetic field dependent magnetization loops, as those shown in Fig.1, were obtained by measuring, at each field, the difference between the TEY intensity at the $M_5$-Eu edge peak, normalized by the intensity below the absorption edge, obtained with two different helicities (combination with polarization and field direction).

SQUID data were collected by using a Quantum Design MPMS3. Magnetization measurements as a function of magnetic field were acquired on both LAO/ETO//STO heterostructure and bare STO substrates that suffered the same heating process mimicking the LAO/ETO growth process to confirm and exclude the absence of ferromagnetic impurities stemming from the substrate itself. Data were corrected for the diamagnetism of the substrate substrating the linear contribution acquired at high magnetic fields.

RESPES Measurements were carried out at a pressure of $5 \times 10^{-10}$ mbar and at base temperature of ≈ 12 K (above the FM Tc) on LAO(5)/ETO(2uc)/STO samples. This technique, employing soft-energy x-rays resonant with the absorption of the relevant ions in the system, enables access to buried systems, and it is, therefore, suitable for the investigation of the q2DEG at the LAO/ETO/STO interface. To reduce any influence of contaminants and to preserve as much as possible the surface of the ex-situ grown samples, they were transferred, just after the deposition, into a sample-carrier vessel filled by inert Ar-gas. However, an exposure (limited in time) to ambient atmosphere was unavoidable for the sample mounting into the experimental station. During the measurements we moved the beam position around the region investigated in order to avoid variations of the surface and interfacial oxygen by photon irradiation, to which LAO/STO samples of certain preparation protocols have demonstrated some sensitivity at low temperature [28]. As matter of fact, by raster scanning the beam on the sample, we did not detect any buildup of the $Ti^{3+}$ spectral weight in the core level x-ray photoemission (XPS) and XAS spectra, nor in the valence band (VB).

DFT+*U* calculations were performed with the Vienna *ab initio* simulation package (VASP) (39, 40) [37][38] with the projector augmented wave (PAW) basis [39][40]. The ionic positions were fully relaxed until the forces were less than 0.001 eV/Å. The generalized gradient approximation was used for the exchange correlation functional in the implementation of Perdew, Burke and Ernzerhof [41] and an on-site effective Hubbard parameter [42], *U* = 4 eV, 7.5 eV and 8 eV were applied on the Ti-*3d*, Eu-*4f* and *La-4f* states, respectively. Analogous to the ETO (001) surface state [12], the choice of *U* for the Eu-*4f* states, and partially for the Ti-*3d* states, is dictated by the need to reproduce the position of the experimental $Eu^{2+}$ peak in the valence band at ~2eV. In general, lower values of *U* for Eu-*4f* only shift the position of the Eu-*4f* band to energies closer to the Fermi level. The results shown here for *U*=4 eV for the Ti-*3d* states give the best agreement with the experimental data, and in particular concerning the position of Eu-*4f* band. Calculations with *U* ranging from 1 eV to 4.5 eV show the main conclusions are not dependent on the choice of *U* for Ti-3*d* states. Further results from the calculations are shown in the Supplementary Materials, from page 24.

**Acknowledgements**

The Authors acknowledge funding from ERA-NET QUANTERA European Union's Horizon H2020 project QUANTOX under Grant Agreement No. 731473, Ministero dell'Istruzione, dell'Università e della Ricerca (MIUR) for the PRIN project TOP-SPIN (Grant No. PRIN 20177SL7HC) and for the PRIN 2010-11 project (Grant No. PRIN 2010-11–OXIDE), the EU COST program CA16218 (Nanocohybri), the German Research Foundation (DFG) within CRC/TRR80 (project number 107745057, subproject C3), CRC1242 (project number 278162697, subproject C02), and Computation time at the Leibniz Rechenzentrum Garching, project pr87ro and supercomputer magnitUDE (DFG grants INST 20876/209-1 FUGG, INST 20876/243-1 FUGG).

Author contributions:

Conceptualization: RDC, DS, GG, MS

Methodology: -

Sample preparation: AS, EDG, FMG, YC, MS

Transport measurements and Analysis: BJ, DS, YC, MDA

RESPES Experiment and Analysis: RDC, VNS, MS, MR, EG, EDG, FMG

ARPES Experiment: MR, GMDL, RDC, BG, NP, MS

XMCD Experiment: and Analysis: CP, MS, GG, GMDL, DP

SQUID Experiment: DP

DFT+U Calculations: MV, RP

Supervision: MS, RDC

Writing: MS, RDC, DS

Writing-review & editing: all the Authors participated to the editing of the paper

# Supplementary Materials

**1. Details of XAS and XMCD analysis and supplementary data.**

1.1 X-ray Absorption Spectroscopy and X-ray magnetic linear and circular dichroism experiment

X-ray Absorption spectra have been acquired using Total Electron Yield (TEY) method. The x-ray magnetic circular dichroism technique allows determining the projection of the magnetic moment associated to a specific ion in the structure along the photon beam direction. The circular polarization of the light carries a moment, which is transferred to the absorbing atom. Selection rules imply different XAS spectra for the two opposite (left and right) circular polarizations if the absorbing ion is characterized by a magnetic moment component along the beam direction.

In the TEY mode we measure a current that is proportional to the photo and Auger electrons created by the photo-absorption process. However, only the photo and Auger electrons created within the electron escape depth reach the surface after elastic and inelastic collisions. Overall, neutrality requires a current of electrons going from the ground to the sample. By measuring this current, we get a measure of the electrons leaving the samples. The electron escape depth is few nm's, and in the particular case of absorption from Ti-$L_{23}$ edge, it is of the order of 1-2 nm.

In the case of transition metal oxides and for soft x-ray absorption processes, i.e. 2p->3d transition, sum rules can be applied to get a quantitative estimate of the effective spin ($m_{Espin}$) and orbital components ($m_{orb}$) of the magnetic moment. However, as discussed in details in ref.(25) of the main text, the application of the sum rules is powerful but needs some care in particular for light transition metals, like titanium.

The Equations to be used are the following (S1):

$$m_{Espin} = -\frac{6\int_{L_3}(I^+ - I^-)dE - 4\int_{L_3+L_2}(I^+ - I^-)dE}{\int_{L_3+L_2}(I^+ + I^-)dE}(10 - n_{3d}) \quad \text{(eq. S1a)}$$

$$m_{orb} = -\frac{4}{3}\frac{\int_{L_3+L_2}(I^+ - I^-)dE}{\int_{L_3+L_2}(I^+ + I^-)dE}(10 - n_{3d}) \quad \text{(eq. S1b)}$$

where $I^+$ and $I^-$ are the XAS spectra acquired with the two circular polarizations, the magnetic dichroism is $XMCD=I^+-I^-$, and $n_{3d}$ is the 3d electronic density. To evaluate eq. S1a it is necessary to separate the $L_3$ and $L_2$ edges in the experimental spectrum and calculate the following integrals:

$$q = \int_{L_3} \left( I^+ - I^- \right) dE$$

$$p = \int_{L_3+L_2} \left( I^+ - I^- \right) dE$$

However, the $L_3$ and $L_2$ contributions are mixed in the Ti XAS spectra, and it is not possible to establish a reliable procedure to separate them. Thus, the evaluation of the spin moment at Ti-edge is affected by an unknown error.

On the other hand, the orbital moment can be calculated safely, since there is no need to separate the integrals at the $L_2$ and $L_3$ edges and the relationship relating the orbital moment to the XMCD is exact.

### 1.2 X-ray Multiple scattering calculations and charge transfer

Simulations including charge transfer effects were done using charge transfer ligand field multiplet theory, implemented in the CTM4XAS software (24). The charge transfer is included in an Anderson impurity type of model where two parameters are necessary to describe the system: the energy difference between the Ti-*3d* and the ligand band (called $\Delta$ here) and the transfer integral which describes the overlap between metal and ligand bands and therefore gives the probability for the electron to be transferred. The $\Delta$ fitting the XAS data is 4 eV. The transfer integrals used were 2 and 1 for $b_1$, $a_1$ ($d_{x2-y2}$, $d_{z2}$) and $b_2$,e ($d_{xy}$, $d_{xz}/d_{yz}$) orbitals, respectively. This choice is commonly used for octahedral or close to octahedral systems and the reasoning behind is that the probability for electron transfer is higher (in this case two times higher) for orbitals pointing in the bond direction, as $d_{x2-y2}$, $d_{z2}$. We used $U_{pd}-U_{dd}=2eV$. The crystal field parameters were $10D_q=2.3$, $D_s=-0.02$, $D_t=-0.002$. A slightly larger value of $10D_q$ compared to the simulation for LAO/STO seems to fit better the energy difference between the double peaks at $L_3$ and $L_2$ edges. The Slater Integrals were reduced by 90% applied on the corrected Hartree-Fock values. This simulation gives a number of 3d holes equals to 9.65. The broadening used were 0.1 eV for the Gaussian broadening, to account for the energy resolution, and different Lorentzian broadening for each of the 4 peaks of the spectrum, namely half width at half maximum of 0.11 eV, 0.34 eV, 0.5 eV, and 0.65 eV for the four XAS peaks (from the first to the last).

Fig. S1 shows the comparison of the simulation with LAO/ETO/STO XAS data. The XAS data agreement is very good, except for the features around 470 eV and 475 eV.

In Fig. S2 we also show simulation for titanium in a $Ti^{3+}$ configuration in D4h symmetry, which resemble the shape of the Constant Initial State (CIS) spectra obtained by integrating the valence band RESPES results shown in Fig.3 of the manuscript. We used the same crystal field splitting parameters employed for $Ti^{4+}$ spectra and CT parameters The charge transfer has the effect of increasing the valleys between the peaks. $\Delta$ =4 eV and $U_{pd}-U_{dd}$=2eV. In the simulations shown below we have also plotted different simulated spectra as function of the 10Dq value from 1.0 eV to 2.2 eV.

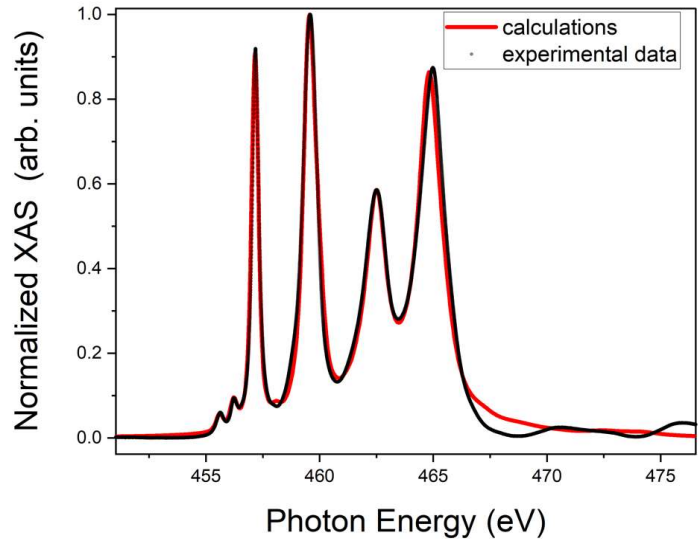

**Fig. S1**: Typical TEY Ti-$L_{2,3}$ edge XAS spectrum on LAO/ETO/STO system in C+ polarization (black circles) and atomic multiplet atomic splitting simulation (red line).

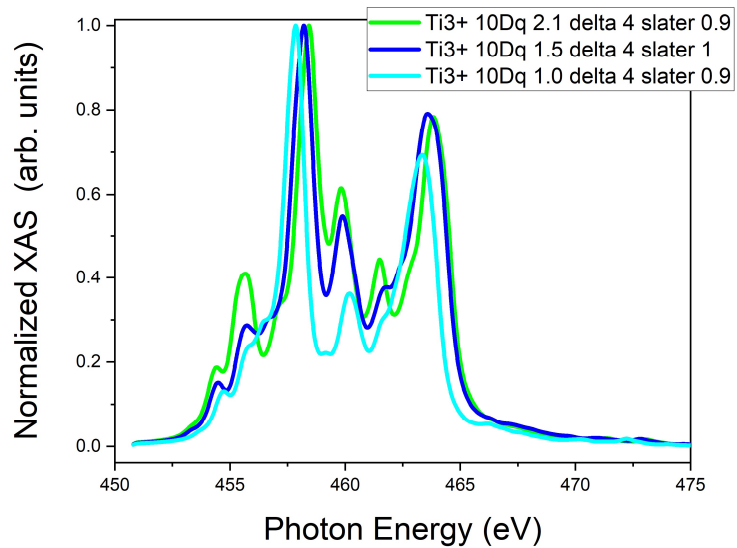

**Fig. S2**: Atomic multiplet splitting simulation assuming $Ti^{3+}$ valence state with varying crystal field (10 Dq) and CT parameters.

## S2. Tight binding fit of the band dispersion

The band-dispersion profiles were analyzed by combining the spectra acquired with different polarizations, and by fitting the Momentum Dispersion Curves (MDC) as function of the energy relative to the Fermi level $E_F$. Finally, the band profiles were compared to the two-dimensional(2D) curvature maps of the raw data (31), as further validation of the fitting process for the band identification.

Circular polarization (C+) *E* vs. $k_x$ band dispersions include features associated to all the bands in the system, namely bands with $3d_{xz}$, $3d_{yz}$ and $3d_{xy}$ orbital characters; dataset recorded in *p-pol* configuration are mainly affected by the $3d_{xz}$ band, while *s-pol* configuration provides information about the $3d_{xy}$ and the $3d_{yz}$ bands. ($d_{xy}$ indicates the light band, while $d_{xz}$ and $d_{yz}$ refer to the heavy bands along the short and the long diameters of the ellipse, respectively).

We assumed for simplicity three independent two-dimensional bands as described by eqs. S2:

$$E_{xy}(k_x, k_y) - E_F = 2V_{xy}(2 - cosk_xa_0 - cosk_ya_0) + E_{0xy} \qquad \text{(eq. S2a)}$$

$$E_{xz}(k_x, k_y) - E_F = 2V_{xz}(1 - cosk_xa_0) + 2V_{yz}(1 - cosk_ya_0) + E_{0xz} \qquad \text{(eq. S2b)}$$

$$E_{xz}(k_x, k_y) - E_F = 2V_{yz}(1 - cosk_xa_0) + 2V_{xz}(1 - cosk_ya_0) + E_{0yz} \qquad \text{(eq. S2c)}$$

where $a_0$ is the lattice constant $V_{xy}$, $V_{xz}$, $V_{yz}$ are the hopping parameters or so-called inner potentials, which determine the band-curvatures and the effective masses, $E_F$ is the Fermi energy, and $E_{0xy}$, $E_{0xz}$, $E_{0yz}$ are the $E - E_F$ value at $k_x$ $k_y$ = 0, i.e. the band bottom of each band ($E_{0xz}$ and $E_{0yz}$ are expected to have the same value inside the experimental error).

In our experimental configuration, recorded maps correspond to *E* vs. $k_x$ curves at $k_y$ = 0, so that eqs. (S2) read as:

$$E_{xy}(k_x) - E_F = 2V_{xy}(1 - cosk_xa_0) + E_{0xy} \qquad \text{(eq. S3a)}$$

$$E_{xz}(k_x) - E_F = 2V_{xz}(1 - cosk_xa_0) + E_{0xz} \qquad \text{(eq. S3b)}$$

$$E_{xy}(k_x) - E_F = 2V_{yz}(1 - cosk_xa_0) + E_{0yz} \qquad \text{(eq. S3c)}$$

The experimental $E$ vs. $k_x$ datasets are obtained by deconvolving MDC intensity profiles of ARPES maps as superposition of Lorentzian contributions. The MDC profile corresponding to a given $\varepsilon = E - E_F$ value is obtained by averaging ARPES intensity values at each $k_x$ over an energy window of 10 meV (centered on the chosen energy). At each given $\varepsilon$ value, the central $k_x$ of each Lorentzian contribution defines the ($k_x$, $\varepsilon$) point to be employed in the tight binding fitting procedure through eqs. (S3).

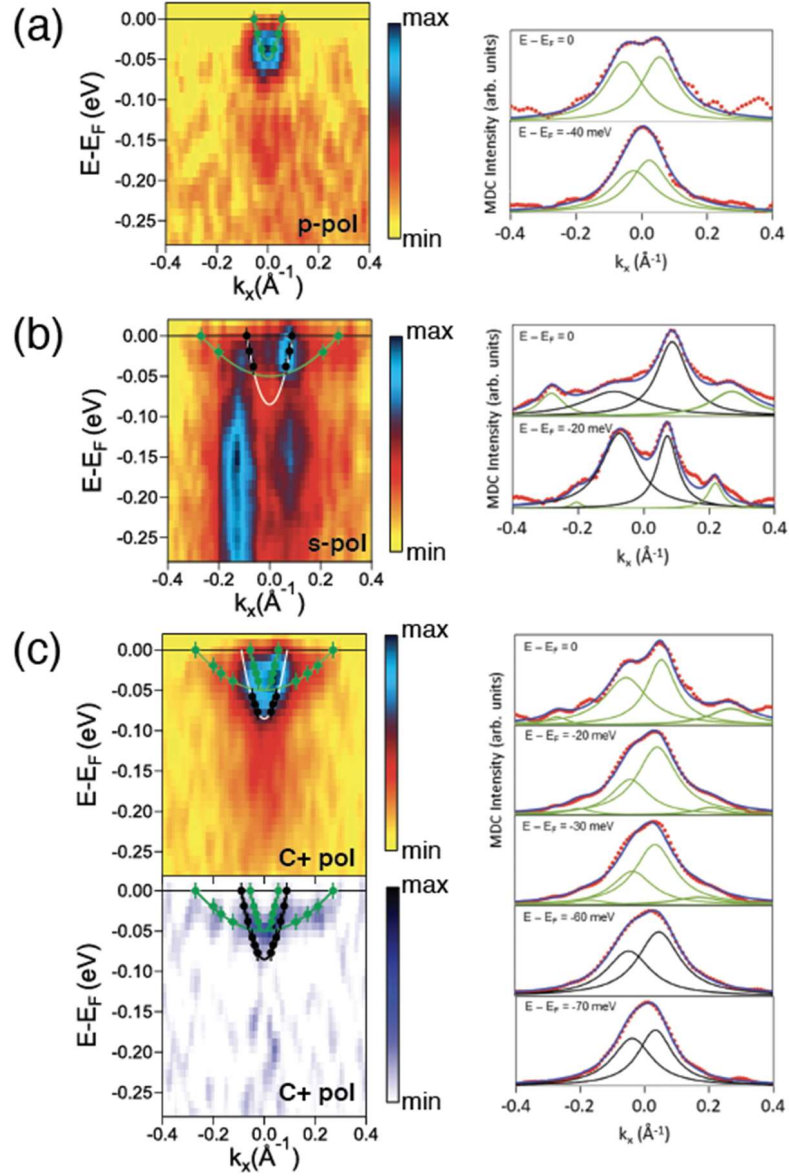

**Fig. S3**: Right panels: Band dispersion ARPES maps recorded with (a) p , (b) s , and (c) C+ polarizations of incident photons in the 2nd Bz. In panel (c) we show also the corresponding 2D curvature map of the C+ data. Left panels: some MDCs profiles at different binding energies (scatter data) extracted from each map and their Lorentzian deconvolutions as described in the text. The experimental points inferred from the deconvolution of the MDCs (black: $d_{xy}$ "light" band; green: $d_{xz}$ and $d_{yz}$ "heavy" bands), as well as the tight binding best fit curves, are overlapped to the ARPES and curvature maps. Only points that could be clearly detected in the deconvolution and assigned were employed. The curves reported on different maps corresponding to the same band were estimated by a joint fit on all the points evaluated on the different maps.

The inferred ($k_x$, $\varepsilon$) points overlapped to ARPES maps (recorded at different polarizations of the incident photons) and to the 2D curvature map in C+ polarization, together with the band profiles as obtained from the fit procedure, are reported in Fig. S3. In the same figure, MDC profiles with their multiple Lorentzian fit are also shown; such Lorentzian deconvolutions provide the experimental points to be employed for the tight binding fit of the bands. For each experimental point, the error bar on the binding energy (vertical scale) is taken as the window of integration used to estimate the MDC curves (10 meV), while the error on the momentum (horizontal scale) is determined from the sensitivity of the corresponding Lorentzian deconvolution.

Finally, the error in the effective masses is determined by propagating the uncertainties associated to the estimation of the Fermi momentum and of the band bottoms of each band, as inferred from the tight binding fits.

In Table S1 we report all the parameters estimated from the fit.

**Table S1**: Fit parameters of the MDC data.

| $V_{xy}$ (eV) | $V_{yz}$ (eV) | $V_{xz}$ (eV) | $E_{0xy}$ (meV) | $E_{0xz}/E_{0yz}$ (meV) | $k_{F\,(xy)}$ (Å$^{-1}$) | $k_{F\,(yz)}$ (Å$^{-1}$) | $k_{F\,(xz)}$ (Å$^{-1}$) |
|---|---|---|---|---|---|---|---|
| 0.70±0.05 | 0.050±0.005 | 1.10±0.05 | -85±5 | -50±5 | 0.09±0.01 | 0.27±0.02 | 0.055±0.005 |

### S3. Supplementary DFT+*U* calculations

In order to illustrate the role of the value of *U* for Ti-*3d* states we show in Fig. S4 a comparison between the band structure calculations in the ground state (FM) of (001) LAO/ETO/STO for Ti-*3d* values of *U*=2eV and *U*=4eV, using the same c(2x2) model described in the main text. One can see that the value of *U* of Ti-*3d* has an effect on the position of the Eu-*4f* peak, which shift to lower energy, around -1.6 eV for *U*=2eV, against the experimental value of -2 eV which is in very good agreement with the simulation done with *U*=4eV. Moreover, the *U*-value also has an effect on the band dispersions and on the FM splitting. The latter is sensibly smaller for *U*=2eV, still it is quite sizeable, thus a FM ground state is a very robust outcome of DFT+*U* calculations. Furthermore, the bottom of the lowest band, having a $3d_{xy}$ orbital character, is shifted to lower energy in the *U*=2eV case, and the $3d_{xz,yz}$ heavy band is shifted to higher energy. Both results give a splitting at Γ which is sensitively different from the experimental result (in both the FM and AFM (not shown) solution).

Thus, from the analysis of the effect of the Ti-*3d U* we conclude that a better agreement with the experimental results is obtained assuming *U*=4eV. On the other hand, the main result, i.e. the FM-ground state of the system, is independent on the choice of the *U*-value.

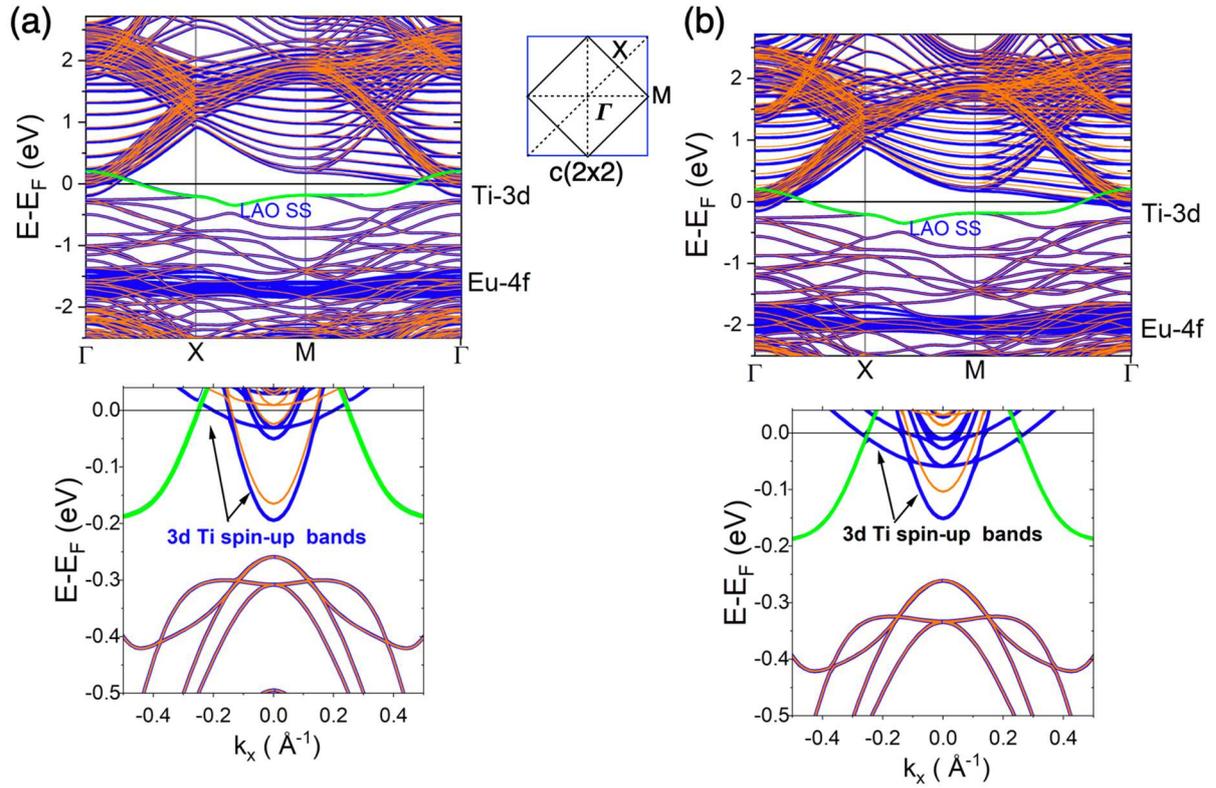

**Fig. S4**: Comparison between DFT+*U* calculations performed with Ti-*3d U* values of (a) 2eV and (b) 4eV.